\documentstyle[12pt]{article}
\def\Vtdabs{\vert V_{td} \vert}

\def\Vtsabs{\vert V_{ts} \vert}

\newcommand{\beq}{\begin{equation}}
\newcommand{\eeq}{\end{equation}}
\newcommand{\ba}{\begin{eqnarray}}
\newcommand{\ea}{\end{eqnarray}}

\def\to{\rightarrow}

\begin{document}
\thispagestyle{empty}
%~\vspace*{-4cm}
\begin{flushright}
DESY 95--106\\
hep-ph/9506248\\
June 1995
\end{flushright}
\vskip0.5cm
\begin{center}
{\Large \bf
Estimates of the Weak Annihilation Contributions to the Decays $B \to \rho
+ \gamma$ and $B \to \omega + \gamma$ }
\vskip1cm
{\large A. Ali} and
 {\large V.M.~Braun} $\footnote { On leave of absence from
St.Petersburg Nuclear
Physics Institute, 188350 Gatchina, Russia.}$ \\
\vskip0.2cm
    Deutsches Elektronen Synchrotron DESY \\
        D--22603 Hamburg (Fed. Rep. Germany)\\
\vskip1cm
{\Large Abstract\\}
\parbox[t]{\textwidth}{
The dominant long-distance contributions
to the exclusive radiative decays $B \to \rho (\omega) +
\gamma$ involve photon emission from the light quarks at large
distances, which cannot be treated perturbatively.
We point out that this emission can be described
 in a theoretically
consistent way as the magnetic excitation of quarks in the QCD vacuum
and estimate the corresponding parity-conserving and parity-violating
amplitudes using the light-cone QCD sum rule approach. These are then
combined with the corresponding short-distance contribution from the
magnetic moment operator in the same approach, derived earlier, to
estimate the decay rates $\Gamma (B \to \rho(\omega) + \gamma)$.
The implications of this result for the determination
of the Cabibbo-Kobayashi-Maskawa (CKM) matrix elements in radiative
 $B$ decays are worked out.} \vskip2cm
{\em Submitted to Physics Letters B}
\end{center}
\newpage
\setcounter{page}{1}
% Decrease texheight (for preprint numbers) again
\textheight 23.0 true cm

%%%%%%%%%%%%%%%%%%%%%% SECTION 1 %%%%%%%%%%%%%%%%%%%%%%%%%%%%%%%%
{\large\bf 1.}\qquad
Exclusive radiative B decays provide a valuable source of information
about the CKM matrix. Assuming that the short-distance (SD) contribution
of the magnetic moment operator is dominating these transitions,
one derives \cite{ABS93}
\beq\label{ratio1}
\frac{\Gamma(B_{u,d} \to \rho \gamma)}{\Gamma(B_{u,d} \to K^* \gamma)}
    = \left( \frac{\Vtdabs}{\Vtsabs}\right)^2 \xi_{u,d} ~,
\label{vckmex}
\eeq
where $\xi_{u,d}$ takes into account $SU(3)$ breaking in the decay form
factors and the (nominal) phase space differences.
With this assumption and isospin invariance one also expects:
\beq\label{ratio2}
\Gamma(B_{u} \to \rho^+ \gamma)=2 ~\Gamma(B_{d} \to \rho^0  \gamma)
    = 2 ~\Gamma (B_{d} \to \omega  \gamma)~.
\eeq
 These relations
have been used
to convert the experimental upper bound on the ratio of the
exclusive radiative $B$ decays \cite{Patterson}:
\beq
 \frac{{\cal B}(B \to \rho(\omega) \gamma)}
      {{\cal B}(B \to K^* \gamma)}< 0.34 ~(90\% \mbox{C.L.}),
\eeq
into a bound $\Vtdabs/\Vtsabs < 0.64-0.76$, depending
on the estimate of the SU(3)-breaking parameters in the SD-piece
\cite{ABS93,Narison,Soares1}.
While this bound is at present not competitive
with the corresponding bound from the CKM unitarity \cite{PDG} and
from the fits of the CKM matrix elements \cite{AL94} which yield
\begin{equation}
 \frac{\Vtdabs}{\Vtsabs} < 0.36\,,
\end{equation}
one anticipates that the increased sensitivity in the
radiative $B$ decay modes
projected for the upgraded CLEO detector and the high luminosity
$B$ factories will
 allow to test
the relationships
{}~(\ref{vckmex}) -- (\ref{ratio2}) quantitatively.

The troublesome point, which
has been raised in a number of papers
\cite{Soni94} -- \cite{Ricciardi}, is
  the possibility of significant
long-distance (LD) contributions to
 radiative $B$ decays from the light quark intermediate states.
Their amplitudes necessarily involve other CKM matrix elements and hence the
simple factorization of the decay rates in terms of the CKM factors
involving $\Vtdabs$ and $\Vtsabs$ no longer holds thereby
 invalidating the relationships ~(\ref{vckmex}) and ~(\ref{ratio2}).
A redeeming feature, however, is that the experiments by testing
{}~(\ref{ratio2})
can determine in a model-independent way the extent to which the radiative
transitions in question are dominated by the SD contributions.
    In general, light quark
$(u,c)$ intermediate states enter in the magnetic moment transitions
through the corresponding Wilson coefficient $C_7^{eff}(\mu)$ (see below),
 and through the transitions induced by the matrix elements of the
 four-Fermion operators.
In fact, the effect of the light quarks to $C_7^{eff}(\mu)$ is
completely negligible as demonstrated through the explicit
calculations \cite{InamiLim,bsgamnew}, being suppressed by powers
$O(m_i/m_W)^2$ and $O(m_i/m_t)^2$ with $m_i=m_u$ or $m_c$ \footnote{ See
also \cite{Ricciardi} for a recent discussion of this point.}.
Estimates of the contribution of the four-Fermion operators
 require a certain non-perturbative
technique and have been worked out using quark models and the vector meson
dominance (VDM) approximation. The
purpose of this letter is to suggest an alternative technique,
which treats the photon emission from the light quarks in a theoretically
consistent and model-independent way. We then combine this treatment
with the light-cone QCD sum rule approach to calculate both the SD and LD
--- parity conserving and parity violating --- amplitudes
in the decays $B_{u,d} \to \rho(\omega) + \gamma$.
We find that the LD contributions are negligible in the
neutral $B$-meson decays $B_d \to \rho(\omega) + \gamma$ but they may
contribute up to $\pm 20 \%$ corrections in the decay rate of the
charged $B$ meson $B_u^\pm \to \rho^\pm + \gamma$. We work out the
modified form of the relations ~(\ref{ratio1}) and ~(\ref{ratio2}), and
 work out the
consequences of this result for the extraction of the CKM parameters
 from exclusive radiative $B$ decays.

\bigskip
%%%%%%%%%%%%%%%%%%%%%% SECTION 2 %%%%%%%%%%%%%%%%%%%%%%%%%%%%%%%%

{\large\bf 2.}\qquad
For subsequent use, we collect the definitions used in this work.
Radiative weak transitions at the $B$-meson scale are governed by the
effective Hamiltonian
\begin{equation}
  {\cal H} =\frac{G_F}{\sqrt{2}}\Big[ V_{ub}V_{ud}^\ast
\Big( C_1 O_1 + C_2 O_2 \Big)
 -V_{tb}V_{td}^\ast C_7^{eff} O_7 +\ldots \Big]\,,
\end{equation}
where we have shown only the contributions which will be important for what
follows.
The $O_1$, $O_2$ are the standard four-Fermion operators
\begin{eqnarray}
 O_1 &=&
       (\bar d_\alpha \Gamma^\mu u_\beta)(\bar u_\beta \Gamma_\mu b_\alpha)\,,
\nonumber\\
 O_2 &=&
       (\bar d_\alpha \Gamma^\mu u_\alpha)(\bar u_\beta \Gamma_\mu b_\beta)\,,
\end{eqnarray}
$\Gamma^\mu =\gamma^\mu (1+\gamma_5)$\footnote{
Our conventions for the $\gamma_5$-matrix
and $\epsilon_{\mu\nu\alpha\beta}$ tensor conform to those in
 \cite{Okun}.}
and $\alpha, \beta$ are color
indices, the $C_1$, $C_2$ are the corresponding coefficients (depending on the
scale $\mu$); $O_7$ is the magnetic moment operator
\begin{equation}
    O_7=\frac{e\,m_b}{8\pi^2} \bar d\sigma_{\mu\nu}
    (1-\gamma_5)F^{\mu\nu} b
\end{equation}
and $F_{\mu\nu}$ is the electromagnetic field
strength tensor, which we take,
for the photon emission with momentum $q$ and polarization
$\varepsilon_\mu^{(\gamma)}$, to be
\begin{equation}\label{plainwave}
   F_{\mu\nu}(x)= i\big(\varepsilon_\nu^{(\gamma)}q_{\mu}
              -\varepsilon_\mu^{(\gamma)}q_{\nu}\big) e^{iqx}\,.
\end{equation}
The coefficient $C_7^{eff}$ includes also the effect of the four-Fermi
operators $O_5$ and $O_6$.
For details and numerical
values of these coefficients, see \cite{Buras,AGM94}.
We concentrate on the $B_u^\pm$ decays,
$B_u^\pm \to \rho^\pm + \gamma$ and take up the neutral $B$ decays
$B_d \to \rho (\omega) + \gamma$ at the end.
The SD contribution to the decay rate involves the matrix element
\begin{equation}
\langle \rho\gamma | O_7|B\rangle =
\frac{e\,m_b}{8\pi^2}(-2i)
\varepsilon^{(\gamma)\mu}
\langle\rho|\bar d\sigma_{\mu\nu}q^\nu (1-\gamma_5)b|B(p)\rangle\,,
\end{equation}
which is parametrized in terms of two invariant form factors
\begin{equation}
 \langle\rho(k)|\bar d\sigma_{\mu\nu}q^\nu (1-\gamma_5)b|B(p)\rangle =
 \Big[\varepsilon_\mu^{(\rho)}(q\cdot p)-
          p_\mu(q\cdot\varepsilon^{(\rho)})\Big] \cdot 2 F_1^{S}(q^2)
  +i\epsilon_{\mu\nu\alpha\beta}\varepsilon^{(\rho)\nu} p^\alpha q^\beta
 \cdot 2 F_2^{S}(q^2)\,.
\end{equation}
Here $p$ and $k=p-q$ are, respectively, the
 $B$-meson and $\rho$-meson momentum and $\varepsilon^{(\rho)}$ is the
polarization vector of the $\rho$ meson.
For the real photon emission the two form factors coincide,
 $  F_1^{S}= F_2^{S}\equiv F^{S}$.
This form factor was calculated in \cite{ABS93} using the light-cone
QCD sum rules, and we shall follow this approach in this paper as well.

Combining the above expressions we get the SD decay amplitude
\begin{equation}\label{Ashort}
{\cal A}_{short} =-\frac{G_F}{\sqrt{2}} V_{tb}V_{td}^\ast C_7
\frac{e\,m_b}{4\pi^2}
\varepsilon^{(\gamma)\mu}\varepsilon^{(\rho)\nu}
\Big\{\epsilon_{\mu\nu\alpha\beta}p^\alpha q^\beta-
i\Big[g_{\mu\nu}(q \cdot p)-p_\mu q_\nu\Big]\Big\}\cdot 2 F^{S}(q^2=0)\,.
\end{equation}

The LD contributions of the four-Fermion operators $O_1$, $O_2$ involve
 two possibilities for the flavour flow, shown schematically in Fig.~1.
In this paper we consider the contribution of the weak annihilation
of valence quarks in the B meson, Fig.~1a,b. It is color-allowed for the
decays of charged B mesons, and is expected to give the
dominant LD contribution in this case.
In the factorization approximation, we write
\begin{equation}\label{factor}
\langle \rho\gamma | O_2|B\rangle =
\langle \rho | \bar d \Gamma_\mu u|0\rangle
\langle \gamma | \bar u \Gamma^\mu b|B\rangle
+
\langle \rho\gamma | \bar d \Gamma_\mu u|0\rangle
\langle 0 | \bar u \Gamma^\mu b|B\rangle ~,
\end{equation}
and make use of the definitions of the decay constants
\begin{eqnarray}
\langle 0 | \bar u \Gamma_\mu b|B\rangle  & =& i p_\mu f_B,
\nonumber\\
\langle \rho | \bar d \Gamma_\mu u|0\rangle &=&
\varepsilon^{(\rho)}_\mu m_\rho f_\rho,
\end{eqnarray}
to reduce the problem at hand to the calculation of simpler form factors
induced by vector and axial-vector currents.
The two terms in (\ref{factor}) correspond in an obvious way to
the contributions of photon emission from the loop containing the
b quark, Fig. ~1a,
 and from the loop of the light quarks, Fig. ~1b, respectively.
The latter subprocess involves an (axial) vector transition with the
the momentum transfer $m_b^2$. According to our estimates, this
contribution is much smaller than the one coming from the first term,
which is dominated by the photon emission from a soft u-quark and
which we shall consider in detail. To that end,
we write down
\begin{eqnarray}\label{FFL}
f_\rho\langle \gamma | \bar u \Gamma_\nu b|B\rangle  &=&
 -e\varepsilon^{(\gamma)\mu}
 \Big\{-i\Big[g_{\mu\nu}(q\cdot p)- p_\mu q_\nu\Big] \cdot 2 F_1^{L}(q^2)
  +\epsilon_{\mu\nu\alpha\beta} p^\alpha q^\beta
 \cdot 2 F_2^{L}(q^2)
\nonumber\\
&&{}+\mbox{\rm contact terms}\Big\}~,
\end{eqnarray}
where
the two form factors describe parity-violating and parity-conserving
amplitudes, respectively.
Note that we have included the factor $f_\rho$ on the l.h.s. of
(\ref{FFL}) to make the form factors dimensionless.
The contact terms indicated on the r.h.s. of (\ref{FFL}) depend on the
gauge chosen for the electromagnetic field and should be omitted as
they are identically cancelled by similar gauge-dependent contact terms
contributing to the second term in (\ref{factor}). Physically, they
correspond to contributions of photon emission from the $W^{\pm}$ boson
and arise in intermediate steps because the factorization
approximation in (\ref{FFL}) introduces charged weak currents.
Adding the contribution of the operator
$O_1$, we obtain the LD amplitude to the decay $B_u \to \rho + \gamma$
in terms of the form factors $F_1^L$ and $F_2^L$,
\begin{eqnarray}\label{Along}
{\cal A}_{long} &=&
-\frac{e\,G_F}{\sqrt{2}} V_{ub}V_{ud}^\ast
\left( C_2+\frac{1}{N_c}C_1\right) m_\rho
\varepsilon^{(\gamma)}_\mu \varepsilon^{(\rho)}_\nu
\nonumber\\&&{}\times
 \Big\{-i\Big[g_{\mu\nu}(q\cdot p)- p_\mu q_\nu\Big] \cdot 2 F_1^{L}(q^2)
  +\epsilon_{\mu\nu\alpha\beta} p^\alpha q^\beta
 \cdot 2 F_2^{L}(q^2)\Big\}\,.
\end{eqnarray}
We now proceed to estimate the form factors defined in
(\ref{FFL}).

%%%%%%%%%%%%%%%%%%%%%% SECTION 3 %%%%%%%%%%%%%%%%%%%%%%%%%%%%%%%%
\bigskip

{\large\bf 3.}\qquad
The principal contribution to the form factors is due to the photon
emission from the u-quark which involves both
contributions of small (of order $1/m_b$) and
large (of order $1/\Lambda_{QCD}$) distances.
A consistent tool to separate them
is provided
by the Operator Product Expansion, and is most easily implemented
using the external field technique \cite{IS}.
To this end we consider quark propagation in the background of the
external electromagnetic field $V_\mu$,
adding to the QCD Lagrangian density an extra term
$\delta{\cal L}=-ej^{em}_\mu(x) V^\mu(x)$ where $j^{em}_\mu$ is the
electromagnetic current.
The photon emission is obtained as the linear term in the expansion of the
relevant Green functions in the external field.
The corresponding formalism has been worked out in detail in \cite{IS,BY,
BBK,BF1} and in this paper we only give a summary of the
results.

To the leading twist accuracy, that is when the quark is propagating
close to the light-cone, all the necessary information about the magnetic
photon
emission is contained in a single (nonperturbative) matrix element
of the gauge-invariant nonlocal operator
$\bar \psi(0) \psi(x)$
at light-like separations $x^2=0$
\cite{BBK,BF1}:
\begin{equation}\label{phigamma}
  \langle \bar \psi (0)\sigma_{\alpha\beta}
{\rm Pexp}\left[i\int_0^1 du\, x_\mu (g A_\mu-e_u V_\mu)(ux)\right]
\psi(x)\rangle_F =
 e_\psi \chi\langle\bar \psi \psi\rangle
\int_0^1 du\,\phi_\gamma(u)F_{\alpha\beta}(ux)\,.
\end{equation}
The $\langle\ldots\rangle_F$ denotes the vacuum expectation value in the
external field $F$. Note that the path-ordered exponential factor
includes both the gluon field $A_\mu$ and the photon field $V_\mu$.
In what follows we drop the gauge factors, assuming the Fock-Schwinger
gauge $x_\mu A_\mu=x_\mu V_\mu =0$.
The normalization is chosen in such a way that
$\int_0^1 du\,\phi_\gamma(u)=1$, $\langle\bar \psi\psi \rangle$ is the
quark condensate and the dimensionful constant $\chi$ has the
physical meaning of being the magnetic susceptibility of the quark
condensate. In the limit of a constant external magnetic field
$F_{\alpha\beta}(x) = {\rm const.}$,
the spins of quarks in the vacuum tend to get oriented along the field
direction, with the average spin being proportional to the quark density
$\langle \bar \psi \psi\rangle$, the applied field $F_{\alpha\beta}$,
 the quark charge $e_\psi$ and the
magnetic susceptibility of the medium \cite{IS}:
\begin{equation}
  \langle \bar \psi \sigma_{\alpha\beta}
\psi\rangle_F =e_\psi \chi\langle\bar \psi \psi\rangle F_{\alpha\beta} ~.
\end{equation}
The value of the magnetic susceptibility is large \cite{chi}
\begin{equation}\label{chi}
  \chi(\mu= 1\,\mbox{\rm GeV}) =-4.4 \,\mbox{\rm GeV}^{-2} ~,
\end{equation}
and because of this the nonperturbative photon emission is
numerically very important.\footnote{ A larger value
$\chi =-5.7 \,\mbox{\rm GeV}^{-2}$ given in the first of refs.\cite{chi}
corresponds to a lower normalization point $\mu=500$ MeV.}

The physical meaning of the response function $\phi_\gamma(u)$
becomes transparent for the particular choice of the external field
as a
plain wave (\ref{plainwave}).  In this case (\ref{phigamma}) can be
rewritten as
\begin{equation}
\int dy \, e^{iqy}
\langle 0|T\{j^{em}_\mu(y) \bar \psi(0) \sigma_{\alpha\beta}
\psi(x)\}|0\rangle
 = e_\psi\chi \langle \psi\psi\rangle
[q_\beta g_{\alpha\mu}-q_\alpha g_{\beta\mu}]
\int_0^1 du\, e^{iuqx}\phi_\gamma(u) ~.
\end{equation}
Comparing this expression with the usual definition of the light-cone
hadron wave functions \cite{CZreport} it is easy to see that
$\phi_\gamma(u)$ defines the distribution amplitude (photon wave function)
describing the photon dissociation into a quark-antiquark
pair with the variable $u$ being just the momentum fraction carried by
the quark.
The shape of this distribution in general depends on the normalization
 scale for the quark fields. However, an accurate analysis shows \cite{BBK}
that unlike meson wave functions \cite{CZreport},
this distribution is close to its asymptotic form
already at low virtualities $\mu\sim 1$ GeV:
\begin{equation}
  \phi_\gamma(u) = 6u(1-u)\,.
\end{equation}

Corrections to (\ref{phigamma}) come from higher-twist contributions
which are formally suppressed by powers of the deviation from the
light-cone $x^2$. For hadron wave functions these can be attributed to
contributions of higher Fock-components in the wave function with a
larger number of constituents. The case of the photon is still specific,
since  additional contributions arise related to operators involving
the photon field instead of the gluon field\footnote{These additional
contributions can also be thought of as contact terms, produced by
operators which vanish on using the equations of motion. The external
field technique avoids all contact terms at the cost of introducing
additional operators.}.
These specific contributions are more important than those involving
quark-antiquark-gluon operators
(whose photon-to-vacuum matrix elements are numerically small,
 see \cite{BBK,BF1} for the list of
all the contributions to twist-4 accuracy and numerical estimates)
 and can be
calculated exactly in terms of the quark condensate.

The most general decomposition of the relevant matrix element to the
twist-4 accuracy involves two new invariant functions (distribution
amplitudes) $g_\gamma^{(1)}(u)$ and $g_\gamma^{(2)}(u)$:
\begin{eqnarray}\label{contact2}
 \langle \bar \psi(0)\sigma_{\alpha\beta}
\psi(x)\rangle_F &=&
 e_\psi \langle\bar \psi\psi\rangle
\int_0^1 du\,F_{\alpha\beta}(ux)
\Big[\chi\phi_\gamma(u)+{x^2}g_\gamma^{(1)}(u)\Big]
\nonumber\\
&+&{}e_\psi \langle\bar \psi\psi\rangle
\int_0^1 du\,g_\gamma^{(2)}(u)\Big[x_\beta x_\eta F_{\alpha\eta}-
x_\alpha x_\eta F_{\beta\eta}-x^2 F_{\alpha\beta}\Big](ux) ~.
\end{eqnarray}
For one particular projection of Lorentz indices the answer
is given in \cite{BBK,BF1}:
\begin{eqnarray}\label{contact1}
  \lefteqn{
\not\!x\sigma_{\alpha\beta}\not\!x
  \langle \bar \psi (0)\sigma_{\alpha\beta}
  \psi(x)\rangle_F =}
\nonumber\\
&=& e_\psi \langle\bar \psi \psi\rangle
\not\!x\sigma_{\alpha\beta}\not\!x
\int_0^1 du\,F_{\alpha\beta}(ux)
\Big[\chi\phi_\gamma(u)-\frac{x^2}{4}(1-u)+\,\mbox{\rm
gluons}+O(x^4)\Big] ~,
\end{eqnarray}
which implies
\begin{equation}\label{eq1}
   g_\gamma^{(1)}(u)-\frac{1}{2}g_\gamma^{(2)}(u) = -\frac{1}{4}(1-u)\,.
\end{equation}
To determine the functions
$g_\gamma^{(1)}(u)$ and $g_\gamma^{(2)}(u)$ separately,
we use the operator identity
\begin{equation}\label{identity1}
 \bar \psi(0)\sigma_{\alpha\beta} \psi(x)\! =\!
\int_0^1 vdv\,\Big[\frac{\partial}{\partial x_\beta}
\bar \psi(0)
\sigma_{\alpha\eta}x_\eta \psi(vx)-(\alpha\leftrightarrow \beta)\Big]
-\epsilon_{\alpha\beta\eta\rho}\int_0^1 vdv\,x_\eta
\frac{\partial}{\partial x_\rho}\bar \psi(0)\gamma_5 \psi(vx)\,.
\end{equation}
 To prove (\ref{identity1}), consider a simpler identity
\begin{equation}\label{identity2}
  \bar \psi(0)\sigma_{\alpha\beta} \psi(x) =\int_0^1 dv\,\Big[
v^2 \frac{d}{dv}\bar \psi(0)\sigma_{\alpha\beta} \psi(vx)+
2v \bar \psi(0)\sigma_{\alpha\beta} \psi(vx)\Big] ~,
\end{equation}
which can easily be checked integrating by parts. Then substitute
$v (d/dv) = x_\xi (\partial/\partial x_\xi)$
and subtract from (\ref{identity2}) the first term on the r.h.s. of
(\ref{identity1}). By simple algebra the difference can be
written as
\begin{equation}
 \int_0^1 v^2 dv\, \bar \psi(0)\Big[x_\rho\partial_\rho \sigma_{\alpha\beta}-
x_\eta\partial_\beta\sigma_{\alpha\eta}
+x_\eta\partial_\alpha\sigma_{\beta\eta}\Big]\psi(vx) ~,
\end{equation}
where $\partial_\eta\equiv \partial/\partial x_\eta$.
 Since for arbitrary Lorentz vectors $x_\alpha,y_\beta$, one has
\begin{equation}
 (x\cdot y)\sigma_{\alpha\beta}-y_\beta\sigma_{\alpha\eta}x_\eta+
y_\alpha\sigma_{\beta\eta}x_\eta = -\epsilon_{\alpha\beta\eta\rho}x_\eta
y_\rho\gamma_5 ~,
\end{equation}
this expression coincides with  the second term on the r.h.s. of
(\ref{identity1}).

To use (\ref{identity1}), note that the vacuum-to-photon transition
matrix element of the second term on the r.h.s. vanishes.
Substituting the general expression (\ref{contact2}) for the
matrix elements in the first term,  one obtains the (exact) relation
\begin{equation}\label{eq2}
 2u^3\int_u^1\frac{dv}{v^4}g_\gamma^{(1)}(u)= g_\gamma^{(2)}(u) ~.
\end{equation}
Solving the system of equations (\ref{eq1}), (\ref{eq2}) we get:
\begin{eqnarray}
    g_\gamma^{(1)}(u) &=&-\frac{1}{8}(1-u)(3-u) ~,
\nonumber\\
    g_\gamma^{(2)}(u) &=&-\frac{1}{4}(1-u)^2 ~,
\end{eqnarray}
which is our final result for the photon emission to twist-4
accuracy. The contributions of the
quark-antiquark-gluon operators can be added using
(\ref{identity1}) and the expressions given in \cite{BF1}. They are
numerically small and are omitted in the numerical analysis given below.
Taking the external field in the form of the plain wave (\ref{plainwave})
we get an equivalent representation:
\begin{eqnarray}
\lefteqn{\int dy \, e^{iqy}
\langle 0|T\{j^{em}_\mu(y) \bar \psi(0) \sigma_{\alpha\beta}\psi
(x)\}|0\rangle=}
\nonumber\\{}\!\!& =&\!\! e_\psi \langle\bar\psi\psi
\rangle[q_\beta g_{\alpha\mu}-q_\alpha g_{\beta\mu}]
\int_0^1\! du\, e^{iuqx}\Big[\chi\phi_\gamma(u)+{x^2}g_\gamma^{(1)}(u)
\Big]
+e_\psi \langle\bar\psi\psi\rangle
\Big\{(q x)[x_\beta g_{\alpha\mu}-x_\alpha g_{\beta\mu}]
\nonumber\\&&{}
+x_\mu[q_\beta x_\alpha-q_\alpha x_\beta]
-x^2[q_\beta g_{\alpha\mu}-q_\alpha g_{\beta\mu}]\Big\}
\int_0^1 du\, e^{iuqx}g_\gamma^{(2)}(u) ~.
\end{eqnarray}

%%%%%%%%%%%%%%%%%%%%%% SECTION 4 %%%%%%%%%%%%%%%%%%%%%%%%%%%%%%%%
\bigskip

{\large\bf 4.}\qquad
The expressions given above are sufficient for the description
of real photon emission in the case that the kinematics of the particular
process ensures that the quark propagates near the light-cone.
To make a quantitative estimate of the form factors, we use the
modification of the QCD sum rule technique, suggested
in \cite{BBK, CZ-B,BF1,ABS93}.
The essence of this approach is to avoid introduction
of model-dependent wave functions of the $B$ meson, replacing it by
a suitable interpolation operator, and using dispersion relations and
duality to pick up the contribution of the $B$ meson.
Following this method, we consider the correlation function
\begin{eqnarray}
T_{\mu\nu}(p,q)&=&\!\!
i^2\int dx\,e^{-ipx}\int dy\,e^{iqy}\langle 0|T\{j_\mu^{em}(y)
\bar u(0)\Gamma_\nu b(0) \bar b(x) i\gamma_5 u(x)\}|0\rangle
\nonumber
\\&\!\!=&\!\!\!\!\!
i[(p\cdot q)g_{\mu\nu}-p_\mu q_\nu] T_1(p^2)+
\epsilon_{\mu\nu\alpha\beta}p_\alpha q_\beta T_2(p^2)+O(p_\nu) +
\mbox{\rm contact terms}\,,
\end{eqnarray}
with fixed $q^2=0$ and $(p-q)^2=m_\rho^2$. The invariant functions
$T_1$ and $T_2$ can be calculated in QCD at large negative $p^2$, in which
case the above formalism to describe photon emission is fully applicable:
higher-twist components in the photon wave functions give rise to
contributions suppressed by powers of $p^2$.\footnote{
By contact terms we indicate contributions which do not vanish
after multiplication by $q_\mu$. They should be calculated and subtracted
to separate the gauge-invariant Lorentz structures. Another possibility
is to use the Fock-Schwinger gauge for the photon field
$x_\mu V_\mu(x)=0$, $V_\mu(0)=0$, in which case the contact terms vanish.}
 On the other hand,
the discontinuity of the functions $T_1(p^2)$, $T_2(p^2)$ at positive $p^2$
is saturated by contributions of meson states:
\begin{equation}
   T_{1,2}(p^2) = \frac{f_B m_B^2}{m_b}\frac{2F^L_{1,2}(0)}{m_B^2-p^2}
+\ldots
\end{equation}
 Making the usual assumption
that the $B$-meson contribution corresponds to the spectral density
integrated over the ``region of duality'', one arrives at the sum
rule for the form factors of interest expressed in terms of integrals
of photon wave functions. In this letter we cannot give a detailed
derivation and refer the reader to \cite{ABS93} for technical discussion.
Repeating the by now standard steps, we arrive at the sum rules:
\begin{eqnarray}\label{SRL1}
 {}\!\!\frac{f_B m_B^2}{m_b f_\rho} 2 F_1^{L}(0)e^{-(m_B^2-m_b^2)/t}\!\!&=&
\int_0^1\frac{du}{u} e^{-(\bar u/u)m_b^2/t}\theta(s_0-m_b^2/u)
\Bigg\{e_u\langle\bar\psi\psi\rangle\Big[\chi\phi_\gamma(u)
\nonumber\\&\hspace*{-1.4cm}-&{}\hspace*{-1.0cm}
\frac{4(m_b^2+ut)}{u^2t^2}g_\gamma^{(1)}(u)\Big]
+\!\frac{3m_b}{4\pi^2}(2u-1)\Big[(e_u-e_b)\bar u
-e_b\ln u\Big]\Bigg\}\,,
\end{eqnarray}
\begin{eqnarray}\label{SRL2}
 {}\!\!\frac{f_B m_B^2}{m_b f_\rho} 2 F_2^{L}(0)e^{-(m_B^2-m_b^2)/t}\!\!&=&
\int_0^1\frac{du}{u} e^{-(\bar u/u)m_b^2/t}\theta(s_0-m_b^2/u)
\Bigg\{\!e_u\langle\bar\psi\psi\rangle\Big[\chi\phi_\gamma(u)
\nonumber\\
&\hspace*{-1.7cm}-&{}\hspace*{-1.2cm}
\frac{4(m_b^2+ut)}{u^2t^2}[g_\gamma^{(1)}(u)-g_\gamma^{(2)}(u)]\Big]
+\!\frac{3m_b}{4\pi^2}\Big[(e_u-e_b)\bar u
-e_b\ln u\Big]\Bigg\}\,,
\end{eqnarray}
where we have introduced a shorthand notation $\bar u=1-u$ and
neglected $m_\rho^2$ compared to $m_b^2,t,$ and $s_0$.
Here $t$ is the Borel parameter
and $s_0$ is the continuum threshold which stands for the cutoff
in the dispersion relation restricting the region of duality for the
$B$ meson.
In both the sum rules, the first term in the curly bracket corresponds
to nonperturbative contributions of photon wave function, and the two
last terms are perturbative contributions of the photon emission from
$u$- and $b$-quark, respectively.
Numerically, the nonperturbative contribution
of leading twist dominates and is the same
 in both the form factors.
Thus, in our approach  the parity-violating and
parity-conserving LD amplitudes are close to each other.
For completeness, we quote the corresponding sum rule from \cite{ABS93}
for the SD form factor:
\begin{eqnarray}\label{SRS}
 \frac{f_B m_B^2}{m_b f_\rho} 2 F^{S}(0)e^{-(m_B^2-m_b^2)/t}&=&
\int_0^1\frac{du}{u} e^{-(\bar u/u)m_b^2/t}\theta(s_0-m_b^2/u)
\Bigg\{m_b\phi_\rho^\perp(u)+u m_\rho g_\perp^{(v)}(u)
\nonumber\\&+&{}
\frac{m_b^2+ut}{4ut}m_\rho g_\perp^{(a)}(u)\Bigg\} ~,
\end{eqnarray}
where $\phi_\rho^\perp(u)$,  $g_\perp^{(v)}(u)$ and $g_\perp^{(a)}(u)$
are the leading-twist $\rho$ meson wave functions, specified in
Sec.~4 of \cite{ABS93}.
It is seen that the structure of sum rules for the SD and LD form factors
is very similar, and many of the uncertainties
(such as the dependence on $f_B$ and $f_\rho$) cancel in their ratio.

Before proceeding to present our numerical results, we would like to
remark that the calculations presented here suggest that the VDM approach
underestimates the LD amplitudes for the photon
emission. The difference between the VMD approach and our method is
two-fold.
 First, the value of the quark magnetic susceptibility
in the VDM approximation $\chi_{VDM} =-2/m_\rho^2$ \cite{BY}
appears to be significantly below the result in (\ref{chi}). Second,
the $\rho$-meson wave function is more ``narrow'' than the photon
wave function, see \cite{CZreport,BBK, ABS93} and since the decay
kinematics picks up contributions with almost the entire momentum carried
by one of the constituents, the $\rho$-meson contribution in VMD is
additionally suppressed. Thus, we expect that the VDM-type
relations between $B\to \rho\gamma$ and $B\to \rho\rho$ amplitudes
receive  significant ( $\sim 50\%$) corrections from contributions of
the excited states.

%%%%%%%%%%%%%%%%%%%%%% SECTION 5 %%%%%%%%%%%%%%%%%%%%%%%%%%%%%%%%
\bigskip

{\large\bf 5.}\qquad
In performing the numerical analysis we conform to the values of the
parameters given in \cite{ABS93}. We note that
all the three sum rules are dominated by the first term in the curly
brackets, and since also the wave functions $\phi_\gamma$ and
$\phi_\rho^\perp$ are similar, the ratio of the SD- and LD- form factors
takes a simple form:
\begin{equation}
    F^L_{1}/F^S \simeq F^L_{2}/F^S \simeq
 \frac{e_u\chi\langle\bar\psi\psi\rangle}{m_b}\simeq 0.01 ~.
\end{equation}
The only important correction to this estimate
 corresponds to the perturbative emission
while nonperturbative twist-4 contributions are in fact negligible.
Assuming the interval $t\sim 5-10$ GeV$^2$
for the  Borel parameter, we obtain:
\begin{equation}\label{result}
 F^L_1/F^S = 0.0125\pm 0.0010\,,\quad F^L_2/F^S = 0.0155\pm 0.0010 ~,
\end{equation}
where the errors correspond to the variation of the
Borel parameter. Including other possible uncertainties, we
expect an accuracy of the ratios in (\ref{result}) to be of order 20\%.
This can be combined with the result of \cite{ABS93}
\begin{equation}
  F^S_{B_u\to\rho\gamma}
=\sqrt{2}F^S_{B_d\to\rho\gamma}
=\sqrt{2}F^S_{B_d\to\omega\gamma}
 =0.24\pm 0.04 ~,
\end{equation}
to extract the absolute values of the LD form factors.
Since the parity-conserving and parity-violating amplitudes turn out
to be close to each other,
the ratio of the LD and the SD contributions reduces to a number
\begin{equation}\label{ratio2p}
{\cal A}_{long}/{\cal A}_{short}=
\frac{4 \pi^2 m_\rho(C_2+C_1/N_c)}{m_b C_7^{eff}}\cdot\frac{F^L}{F^S}
\cdot\frac{V_{ub}V_{ud}^\ast}{V_{tb}V_{td}^\ast} ~.
\end{equation}
Using $C_2=1.10$, $C_1=-0.235$, $C_7^{eff}=-0.306$ (at the scale $\mu=5$ GeV)
\cite{Buras,AGM94}, we get
\begin{equation}\label{result2}
R_{L/S}^{B_u\to\rho\gamma} \equiv
 \frac{4 \pi^2 m_\rho(C_2+C_1/N_c)}{m_b C_7^{eff}}
\cdot\frac{F^L}{F^S}=-0.30\pm 0.07 ~.
\end{equation}
Since the Wilson coefficients are scale-dependent, this ratio is in
general also scale-dependent (unless cancelled by a compensating dependence
in the form factors.)
 Varying the scale $\mu$ of the coefficients in
the range $m_b/2 \leq \mu \leq 2 m_b$, an additional dependence of $\pm
10\%$ is introduced in $R_{L/S}^{B_u\to\rho\gamma}$, which,
however, is smaller than
the error given above. To get a ball-park estimate of the ratio
${\cal A}_{long}/{\cal A}_{short}$, we take the central values of
the CKM matrix elements,
 $V_{ud}=0.9744\pm 0.0010$ \cite{PDG},
$|V_{td}|=(1.0\pm 0.045)\cdot 10^{-2}$,
$|V_{cb}|=0.041\pm 0.004$ and $|V_{ub}/V_{cb}|=0.08\pm 0.02$ \cite{AL94},
 yielding,
\begin{equation}
|{\cal A}_{long}/{\cal A}_{short}|^{B_u\to\rho\gamma}
= |R_{L/S}^{B_u\to\rho\gamma}|
\frac{|V_{ub}V_{ud}|}{|V_{td}V_{bt}|} \simeq 10\% ~.
\end{equation}
A quantitative analysis of the relative decay rates
 in terms of the CKM parameters is presented
below.

The analogous LD-contributions to the neutral $B$ decays
$B_d\to\rho\gamma $ and $B_d\to\omega\gamma $ are much smaller, a point
that has also been noted in the context of the VMD and quark model
based estimates. In our approach,
 the corresponding form factors for the decays
$B_d \to \rho(\omega)  \gamma$ are obtained from
the ones for the decay $B_u\to\rho \gamma$ reported above by the
replacement $e_u\to e_d$, which gives the factor $-1/2$; in addition,
and more importantly, the
LD-contribution to the neutral $B$ decays
is colour-suppressed, which reflects itself
through the replacement of the factor
$a_1\equiv C_2+C_1/N_c$ in (\ref{ratio2}) by
$a_2\equiv C_1+C_2/N_c$. This yields for the ratio
\begin{equation}
\frac{R_{L/S}^{B_d\to\rho\gamma}}{R_{L/S}^{B_u\to\rho\gamma}}=
\frac{e_d a_2}{e_u a_1} \simeq -0.13 \pm 0.05 ,
\end{equation}
where the numbers are based on using
$a_2/a_1 = 0.27 \pm 0.10$ \cite{BHP93}. This would then yield at most
$R_{L/S}^{B_d\to\rho\gamma} \simeq R_{L/S}^{B_d\to\omega\gamma}=0.05$,
which in turn gives
 ${\cal A}_{long}^{B_d\to\rho\gamma}/{\cal A}_{short}^{B_d\to\rho\gamma}
\leq 0.02$.
We note that the LD-contribution just discussed is of the
same order as the one expected from the diagrams in Fig.~1b,c which we do not
consider in this paper. In what follows we shall therefore neglect
the LD-contribution to the neutral $B$ decays. Restricting ourselves to the
colour-allowed LD-contributions only, the relation ~(\ref{ratio2}) gets
modified to:
\begin{eqnarray}\label{ratio5}
\lefteqn{\frac{\Gamma(B_u\to \rho\gamma)}{2\Gamma(B_d\to \rho\gamma)}
=\frac{\Gamma(B_u\to \rho\gamma)}{2\Gamma(B_d\to \omega\gamma)}
 =\left|1+R_{L/S}^{B_u\to\rho\gamma}
\frac{V_{ub}V_{ud}^\ast}{V_{tb}V_{td}^\ast}\right|^2 =
}
\nonumber\\&&{}
=1+2\cdot R_{L/S} V_{ud}\frac{\rho(1-\rho)-\eta^2}{(1-\rho)^2+\eta^2}
+(R_{L/S})^2 V_{ud}^2\frac{\rho^2+\eta^2}{(1-\rho)^2+\eta^2}\,.
\end{eqnarray}
where $R_{L/S}\equiv R_{L/S}^{B_u\to\rho\gamma}$ and $\rho$, $\eta$
are the Wolfenstein parameters \cite{Wolfenstein}.
The ratio
$\Gamma(B_u\to \rho\gamma)/2\Gamma(B_d\to \rho\gamma)
(=\Gamma(B_u\to \rho\gamma)/2\Gamma(B_d\to \omega\gamma))$
is shown in Fig. ~2
 as a function of the parameter $\rho$, with
 $\eta= 0.2, ~0.3$  and $0.4$.
This suggests that
a measurement of this ratio would constrain the Wolfenstein parameters
$(\rho, \eta)$, with the dependence on $\rho$ more marked. In particular,
a negative value of $\rho$ leads to a
 constructive interference in
$B_u\to\rho\gamma$ decays, while large positive values of $\rho$ give
a destructive interference. This behaviour is in qualitative agreement
with what has been also pointed out in \cite{Cheng}.

The ratio of the CKM-suppressed and CKM-allowed
 decay rates in (\ref{ratio1}) for charged $B$ mesons
likewise gets modified due to the LD contributions:
\begin{eqnarray}\label{ratio3}
\lefteqn{\frac{\Gamma(B_u\to \rho\gamma)}{\Gamma(B\to K^*\gamma)}
= \xi_u \lambda^2[(1-\rho)^2+\eta^2]
}
\nonumber\\&&{}
\times\Bigg\{
1+2\cdot R_{L/S} V_{ud}\frac{\rho(1-\rho)-\eta^2}{(1-\rho)^2+\eta^2}
+(R_{L/S})^2 V_{ud}^2\frac{\rho^2+\eta^2}{(1-\rho)^2+\eta^2}\Bigg\}\,,
\end{eqnarray}
where $\lambda=0.2205$.
 Using the central value from the estimate
 $\xi_u=0.59\pm 0.08$ \cite{ABS93} we show in Fig.~3a the
ratio (\ref{ratio3}) as a function of $\rho$ for
 $\eta=0.2, ~0.3$ and $0.4$. It is seen that the dependence of this ratio
is rather weak on the parameter $\eta$ but it depends on
 $\rho$ rather sensitively. To show the effect of the LD
contribution,
 we compare in Fig.~3b
 the predictions for the ratio ~(\ref{ratio3}) as a function of $\rho$
 with and without the LD-contribution, fixing
$\eta=0.3$, the central value for this parameter obtained from the CKM
fits \cite{AL94}.
 It is seen that
the effect of the LD contributions on this ratio is small, and is
comparable to $\sim 15\%$ uncertainty in the normalization
due to the $SU(3)$-breaking effects in the form factors.
Neglecting the colour-suppressed LD contributions, as argued above,
the ratio of the decay rates for neutral $B$ meson is not effected
and to a good approximation the SD-dominated result
\begin{equation}
\frac{\Gamma(B_d\to \rho\gamma,\omega\gamma)}{\Gamma(B\to K^*\gamma)}
 = \xi_d\lambda^2 [(1-\rho)^2+\eta^2]
\end{equation}
still holds. Finally, combining the estimates presented here and
 in \cite{ABS93} for the form factors and restricting the Wolfenstein
parameters in the range $-0.4 \leq \rho \leq 0.4$ and $ 0.2 \leq \eta
\leq 0.4$, as suggested by the CKM-fits \cite{AL94}, we give the
following range for the absolute branching ratios:
\begin{eqnarray}\label{ratio4}
{\cal B}(B_u\to \rho\gamma)
&=& (1.9 \pm 1.6) \times 10^{-6} ~,
\nonumber\\
{\cal B}(B_d\to \rho\gamma) &\simeq& {\cal B}(B_d \to \omega \gamma)
= (0.85 \pm 0.65) \times 10^{-6} ~,
\end{eqnarray}
where we have used the experimental value for the branching ratio
${\cal B} (B \to K^* + \gamma) =(4.5 \pm 1.5 \pm 0.9) \times 10^{-5}$
\cite{CLEOrare1},
adding the errors in quadrature. The large error reflects the poor
knowledge of the CKM matrix elements and hence experimental determination
of these branching ratios will put rather stringent constraints on the
parameters $\rho$ and $\eta$. Similar constraints would also follow from the
measurement of the inclusive radiative decays $B \to X_d + \gamma$
\cite{ag2}, for which a branching ratio ${\cal B}(B \to X_d + \gamma)=
(1.0 \pm 0.8) \times 10^{-5}$ is estimated using the measured inclusive
rate ${\cal B}(B \to X_s + \gamma)= (2.32 \pm 0.57 \pm 0.35) \times 10^{-4}$
\cite{CLEOrare2}.

To summarize, we have presented a QCD sum rule-based
calculation of the contribution of the
weak annihilation to exclusive radiative rare B decays, which is expected to
dominate the LD contribution for the decays of charged $B$ mesons,
$ B_u^\pm \to \rho^\pm + \gamma$.
The numerical
 effect of the weak annihilation amplitude depends on the value of the
CKM-Wolfenstein parameter $\rho$,
and is typically estimated in the 10\% range for the presently allowed
values of the parameters $\rho$ and $\eta$. The corresponding LD
contributions in the neutral $B$-meson decays $B_d \to \rho (\omega) \gamma$
are found to be insignificant.
The interference of LD and SD contributions provides a possibility to
measure the sign of the
Wolfenstein parameter $\rho$ from the ratio (\ref{ratio5}).
The branching ratios
 ${\cal B}(B_u \to \rho \gamma)$ and ${\cal B} (B_d \to \rho(\omega)
\gamma)$ likewise are sensitive to the sign and magnitude of $\rho$,
increasing with large negative values of $\rho$.
Our investigations strengthen the expectations that
exclusive radiative rare $B$ decays, very much like their inclusive
counterparts, are dominated by SD contributions, which in the context of
the standard model implies that these decays are
valuable in quantifying the CKM parameters.

Acknowledgements:
  One of us (A.A.) would like to thank Amarjit Soni and Giulia
Ricciardi for vigorous
discussions on radiative $B$ decays.
 The other (V.M.B.)
would like to thank  Alexander
Khodzhamirian for a correspondence and helpful discussions.

%%%%%%%%%%%%%%%%%%%%% REFERENCES %%%%%%%%%%%%%%%%%%%%%%%%%%%%%%%%

\noindent
{\bf Figure Captions}\\

\noindent
Figure 1: Weak annihilation contributions in $B_u \to \rho  \gamma$
involving the operators $O_1$ and $O_2$ denoted by $\bigotimes$ with the
photon emission from a) the loop containing the $b$ quark, b) the loop
containing the light quark, and c) the tadpole which contributes only
with additional gluonic corrections.\\

\noindent
Figure 2: Ratio of the neutral and charged $B$-decay rates
 $\Gamma (B_u \to \rho \gamma)/2\Gamma (B_d \to \rho \gamma)$ as a function
of the Wolfenstein parameter $\rho$, with $\eta =0.2$ (short-dashed curve),
$\eta =0.3$ (solid curve), and $\eta =0.4$ (long-dashed curve).\\

\noindent
Figure 3: Ratio of the CKM-suppressed and CKM-allowed radiative $B$-decay
rates\\
$\Gamma (B_u \to \rho \gamma)/\Gamma (B \to K^* \gamma)$ (with $B=B_u$ or
$B_d$) as a function of the Wolfenstein parameter $\rho$,
a) with $\eta =0.2$ (short-dashed curve), $\eta =0.3$ (solid curve), and
$\eta =0.4$ (long-dashed curve);
b) comparison of the result obtained by neglecting the LD contribution
(dashed curve) with the one including the LD contribution (solid curve),
both evaluated with $\eta =0.3$.


\newpage
\begin{thebibliography}{12}
%\newcommand{\artref}[4]{{\sc #1}, #2 \underline{#3} #4}
%\newcommand{\bookref}[2]{{\sc #1}, #2}

\bibitem{ABS93}
A. Ali, V.M. Braun, and H. Simma, Z. Phys. {\bf C63} (1994) 437.

\bibitem{Patterson}
 R. Patterson,
 in Proc. of the XXVII
Int. Conf. on High Energy Physics,
Glasgow, Scotland, 1994, eds. P.J. Bussey and I.G. Knowles (IOP Publ. Ltd.,
Bristol, 1995).

\bibitem{Narison}
S. Narison, Phys. Lett. {\bf B327} (1994) 354.

\bibitem{Soares1}
J.M. Soares, Phys. Rev. {\bf D49} (1994) 283.

\bibitem{PDG}
L. Montanet et ~al. (Particle Data Group), Phys. Rev.
 {\bf D50}  (1994) 1173.

\bibitem{AL94} A. Ali and D. London, Z. Phys. {\bf C65} (1995) 431.


\bibitem{Soni94}
D. Atwood, B. Blok, and A. Soni, TECHNION-PH-94-11 (1994)
 (hep-ph-9408373).
\bibitem{Deshpande}
N.G. Deshpande, X.-G. He, and J. Trampetic, Preprint OITS-564(1994).
\bibitem{Cheng}
H.-Y. Cheng, Preprint IP-ASTP-23-94 (hep-ph-9411330) (1994).
\bibitem{Soares2}
J.M. Soares, Preprint TRI-PP-95-6 (1995) (hep-ph-9503285).
\bibitem{Milana}
J. Milana, Preprint hep-ph-9503376 (1995).
\bibitem{EIM95}
G. Eilam, A. Ioannissian, and R.R. Mendel, Preprint TECHNION-PH-95-4
      (hep-ph-9505222).
\bibitem{Ricciardi}
G. Ricciardi, Preprint HUTP-94/A037, DSF-T-95/2 (1995).

\bibitem{InamiLim}
     T. Inami and C.S. Lim,
     Progr. Theor. Phys. {\bf 65} (1981) 297.

\bibitem{bsgamnew}
M. Ciuchini et al., Phys. Lett. {\bf B316} (1993) 127; Nucl. Phys.
 {\bf B415} (1994) 403;\\
 G. Cella et al., Phys. Lett.  {\bf B325} (1994) 227.

\bibitem{Buras}
A.J. Buras et al., Nucl. Phys. {\bf B424} (1994) 374.

\bibitem{AGM94}
A. Ali, G. Giudice and T. Mannel, CERN-TH.7346/94.
\
\bibitem{IS}
B.L. Ioffe and A.V. Smilga, Nucl. Phys. {\bf B232} (1984) 109.

\bibitem{BY}
I.I. Balitsky and A.V. Yung, Phys. Lett. {\bf B129} (1983) 328.

\bibitem{BBK}
I.I.~Balitsky, V.M.~Braun and A.V.~Kolesnichenko,
 Sov. J. Nucl. Phys.   {\bf 44 } (1986) 1028;
 Nucl. Phys.  {\bf  B312}  (1989) 509.

\bibitem{Okun}
"Leptons and Quarks", L.B. Okun, (North-Holland Publishers, 1984).
\bibitem{BF1}
V.M. Braun and I.E. Filyanov,
 Z. Phys.  {\bf C44}  (1989) 157.

\bibitem{CZreport}
V.L. Chernyak and A.R. Zhitnitsky
  Phys. Rep.  {\bf {112}} (1984) 173.

\bibitem{chi}
V.M. Belyaev and Ya.I. Kogan, Yad. Fiz. {\bf 40} (1984) 1035;\\
I.I. Balitsky, A.V. Kolesnichenko and A.V. Yung,
Yad. Fiz. {\bf 41} (1985) 282.

\bibitem{CZ-B}
V.L.~Chernyak and I.R.~Zhitnitsky,
 Nucl. Phys.  {\bf B345 } (1990) 137.

\bibitem{Wolfenstein}
        L. Wolfenstein, Phys. Rev. Lett. {\bf 51} (1983) 1945.

\bibitem{BHP93}
T. Browder, K. HonScheid, and S. Playfer, in: {\it B Decays}, 2nd. Edition,
ed. S. Stone (World Scientific, Singapore, 1994).

\bibitem{CLEOrare1}
 R. Ammar et al. (CLEO Collaboration), Phys. Rev. Lett. {\bf 71} (1993) 674.

\bibitem{ag2}  A. Ali and C. Greub ,
               Phys. Lett. {\bf B287} (1992) 191.

\bibitem{CLEOrare2}
M.S. Alam et al. (CLEO Collaboration), Phys. Rev. Lett. {\bf 74} (1995) 2885.

\end{thebibliography}
\end{document}